\documentclass[twocolumn,pra,superscriptaddress]{revtex4-2}
\usepackage[english]{babel}
\usepackage[utf8]{inputenc}
\usepackage[colorinlistoftodos, color=green!40, prependcaption]{todonotes}
\usepackage{color}
\usepackage{amsthm}
\usepackage{mathtools}
\usepackage{amsmath,amssymb}
\usepackage{physics} 
\usepackage{xcolor}
\usepackage{graphicx}
\usepackage{soul}
\usepackage{qcircuit}
\usepackage{bm}
\usepackage{bbm}
\usepackage{mathbbol}
\usepackage{microtype}

\usepackage{float}

\usepackage{bbold}
\usepackage[normalem]{ulem}

\begin{document}

\title{Mitigating Quantum Gate Errors for Variational Eigensolvers Using Hardware-Inspired Zero-Noise Extrapolation}

\author{Alexey Uvarov}
    \affiliation{Skolkovo Institute of Science and Technology,     Moscow, 121205, Russia}
    
\author{Daniil Rabinovich}
\affiliation{Skolkovo Institute of Science and Technology,     Moscow, 121205, Russia}
\affiliation{Moscow Institute of Physics and Technology, Dolgoprudny, 141700, Russia}

\author{Olga Lakhmanskaya}
\affiliation{The Russian Quantum Center,     Moscow, 121205, Russia}

\author{Kirill Lakhmanskiy}
\affiliation{The Russian Quantum Center,     Moscow, 121205, Russia}      
    
\author{Jacob~Biamonte}
\affiliation{Yanqi Lake Beijing Institute of Mathematical Sciences and Applications, Yanqi Island, Huairou District, Beijing, 101408, China}

\author{Soumik Adhikary}
\affiliation{Skolkovo Institute of Science and Technology,     Moscow, 121205, Russia}

\date{\today}

\begin{abstract}
Variational quantum algorithms have emerged as a cornerstone of contemporary quantum algorithms research. Practical implementations of these algorithms, despite offering certain levels of robustness against systematic errors, show a decline in performance due to the presence of stochastic errors and limited coherence time. In this work, we develop a recipe for mitigating quantum gate errors using zero-noise extrapolation. 
We introduce an experimentally amenable method to control error strength in the circuit. 
We utilize the fact that gate errors in a physical quantum device are distributed inhomogeneously over different qubits and qubit pairs. 
As a result, one can achieve different circuit error sums based on the manner in which abstract qubits in the circuit are mapped to a physical device. 
We apply the proposed protocol to variational quantum algorithms and find that the estimated energy is approximately linear with respect to the circuit error sum (CES). Consequently, a linear fit through the energy-CES data, when extrapolated to zero CES, can approximate the energy estimated by a noiseless variational algorithm. 
We demonstrate this numerically and investigate the applicability range of the technique.
\end{abstract}

\maketitle

\section{Introduction}

Variational algorithms are designed to operate within the practical limitations of near-term quantum computers which are inherently noisy. Such algorithms are known to partially alleviate certain systematic limitations of near-term devices, such as variability in pulse timing and limited coherence times \cite{harrigan2021quantum,pagano2019quantum, guerreschi2019qaoa,butko2020understanding}. This is accomplished by the use of a short depth parameterized circuit where the parameters are trained using a quantum-to-classical feedback loop. Equipped with these practical advantages, variational algorithms have found several applications, such as quantum approximate optimization (QAOA) \cite{niu2019optimizing,Farhi2014,lloyd2018quantum,morales2020universality,Farhi2016,Akshay2020,farhi_quantum_2022,Wauters2020,Akshay2021parameter,Campos2021,Rabinovich2022progress,Akshay2022circuit,Rabinovich2022ion}, variational eigensolvers (VQE)  \cite{peruzzo_variational_2014, cao_quantum_2019, Mc-ardle2020, Wecker2015, Bauer2016} and quantum assisted machine learning \cite{biamonte2017, Schuld2019, Adhikary2020, Adhikary2021, Havlivcek2019}. 

Variational algorithms are susceptible to stochastic noise~\cite{fontana_evaluating_2021}. For example, noise can induce barren plateaus in the training landscape, thereby making optimization difficult \cite{wang_noise-induced_2020}. Although small amounts of noise can remove local minima in the cost landscape~\cite{Campos2021}, in general the presence of stochastic noise results in the worsening of the final outcome of the algorithm. This necessitates quantum error mitigation \cite{cai2022}. While there are several methods to mitigate errors in a quantum circuit~\cite{bravyi_mitigating_2021, czarnik_error_2021, endo_practical_2018, lowe_unified_2021, maciejewski_mitigation_2020, pokharel_demonstration_2018, temme_error_2017, viola_dynamical_1998, zhang_error-mitigated_2020}, for the purpose of this paper we focus on  zero-noise extrapolation (ZNE)~\cite{temme_error_2017, li2017_a, kandala_error_2019}.  In ZNE, an algorithm is executed at different levels of noise, in order to establish a dependence between the output of the algorithm and the noise strength. The dependence is then extrapolated to the zero noise limit giving an approximation of the output of the algorithm in noiseless conditions. 

To execute ZNE, one must be able to scale the strength of the noise in a controllable manner. In this paper we introduce an experimentally amenable method to scale the circuit noise strength. The rationale behind our approach comes from the realization that qubits are not made equal; two-qubit gates acting on different pairs of qubits can have different fidelities \cite{Pogorelov2021, Wright2019}. Therefore the level of noise encountered while executing a circuit is determined by how the abstract qubits in the circuit are mapped to their physical counterparts. Thus by choosing different abstract-to-physical qubit mappings one can control how noise changes in a circuit. Applying this approach to the VQE algorithm, we show that the energy estimated by a noisy VQE is approximately linear with respect to the total circuit error sum. We extrapolate this linear trend to the zero noise limit and show that ZNE recovers the noise-free energy estimation with high accuracy. In fact, we establish that for certain types of variational circuits it is guaranteed that ZNE would recover the exact energy as estimated by noise-free VQE. 
In addition, we investigate the behavior of ZNE with increasing strength of the noise and observe that the extrapolation error grows quite modestly, following an approximately linear scaling. Finally, we show that the performance of the proposed ZNE protocol is better or comparable to the existing ZNE techniques, while not suffering from their drawbacks.

The manuscript is structured as follows. In Section~\ref{sec:vqe}, we briefly recall the variational quantum eigensolver algorithm. In Section \ref{sec:perturb_theory}, we describe the behavior of the energy estimated by VQE in presence of small errors. In Section~\ref{sec:explain_fit}, we propose our method of scaling noise for ZNE and apply it to VQE. In Section~\ref{sec:numerics} we present the results of numerical experiments supporting our proposal. Section~\ref{sec:conclude} concludes the paper.

\section{Variational quantum eigensolver}
\label{sec:vqe}

The variational quantum eigensolver is purpose built to approximate the ground state and the ground state energy of a given Hamiltonian $H$ of $n$ qubits. The algorithm starts with the preparation of a so-called variational state $\ket{\psi(\boldsymbol{\theta})} = U(\boldsymbol{\theta}) \ket{0}^{\otimes n} = \Big(\prod_{j=1}^p U_j (\theta_j) \Big)\ket{0}^{\otimes n}$, where $U(\boldsymbol{\theta})$ is a variational ansatz and $\boldsymbol{\theta} \in [0, 2\pi)^{\times p}$ are tunable parameters. Subsequently, local measurements are performed on the variational state to recover the expectation values of Pauli strings $P_{\boldsymbol{\alpha}}$,
which are then classically processed to construct the energy function of the given Hamiltonian $\braket{\psi(\boldsymbol{\theta})|H}{\psi(\boldsymbol{\theta})} = \sum_{\boldsymbol{\alpha}} h_{\boldsymbol{\alpha}} \bra{\psi(\boldsymbol{\theta})} P_{\boldsymbol{\alpha}} \ket{\psi(\boldsymbol{\theta})}$. Here $h_{\boldsymbol{\alpha}} \in \mathbb{R}$ and $P_{\boldsymbol{\alpha}} = \otimes_{j=1}^n \sigma_{\alpha_j}$ with $\alpha_j \in \{0,1,2,3\}$. In this step we make use of the fact that any Hamiltonian admits an expansion in the basis of Pauli strings. Finally the energy function is minimized using a classical co-processor which outputs:
\begin{eqnarray}
    &\boldsymbol{\theta}^\star \in \arg \min_{ \boldsymbol{\theta}} \braket{\psi(\boldsymbol{\theta})|H}{\psi(\boldsymbol{\theta})} \\
   \nonumber \\
    &E^\star = \min_{ \boldsymbol{\theta}} \braket{\psi(\boldsymbol{\theta})|H}{\psi(\boldsymbol{\theta})} \\
    \nonumber \\
    &\ket{\psi(\boldsymbol{\theta}^\star)} = U(\boldsymbol{\theta}^\star) \ket{0}^{\otimes n}.
\end{eqnarray}
 Here, $\ket{\psi(\boldsymbol{\theta}^\star)}$ is a $p$-depth approximate ground state of $H$. The proximity of $\ket{\psi(\boldsymbol{\theta}^\star)}$ to the true ground state of $H$ typically can not be determined a priori, based only on the minimization of the energy function. Nevertheless, one can establish bounds on their overlap following the stability lemma \cite{biamonte_universal_2021}. 

Over time several improvements in VQE have been reported. For example, there is a number of techniques that do not fix the ansatz circuit $U(\boldsymbol{\theta})$ in advance, but instead construct it during the optimization \cite{grimsley_adaptive_2019,tang_qubit-adapt-vqe_2021,sim_adaptive_2021,bilkis2021,sapova_variational_2021}. Other techniques include efficient estimation of the gradient of the energy function to enable gradient descent~\cite{mitarai_quantum_2018,schuld_evaluating_2019,kyriienko_generalized_2021-1}, grouping terms to lower the number of measurements~\cite{verteletskyi_measurement_2020}, tailoring the ansatz circuit to the restrictions of the problem~\cite{barkoutsos_quantum_2018}, and many more. An extensive general review of current usage of VQE can be found in Ref.~\cite{tilly_variational_2022}, while Refs.~\cite{cao_quantum_2019,elfving_how_2020} discuss the variational algorithms specifically in the context of quantum chemistry.

In addition to the theoretical advances, several recent experiments have demonstrated VQE implementation on physical hardware \cite{kandala2017hardware, shaydulin2023qaoa, pelofske2023high, sack2024large}. Nevertheless, the results clearly indicate diminishing algorithmic performances due to the presence of hardware noise. Motivated by such observations, error mitigation has been considered as a mean to boost the VQE performance. See Section 8 in \cite{tilly_variational_2022} for current error mitigation techniques applied to VQE.

\section{Hardware inspired ZNE}
\label{sec:results}

Consider a variational circuit comprising of single-qubit gates and two-qubit gates arranged in $d$ structurally identical layers. We denote a single-qubit gate belonging to the layer $l$, applied to the qubit $j$ as $G^l_j$ and a two-qubit gate belonging to the layer $l$, applied to the qubit pair $(j,k)$ as $G^l_{jk}$. We define $T$ to be the set of qubit pairs to which the two-qubit gates are applied. The set $T$ is typically determined by the ansatz that is being considered. We use this circuit to variationally minimise a problem Hamiltonian $H$.

Here we assume that the single-qubit gates are noiseless. This is in line with many experimental observations that single-qubit gates have very low infidelities which do not influence the performance of our algorithm \cite{Akerman_2015, Ballance2016, Bermudez2017, Levine2018, Bruzewicz2019}. The two-qubit gates on the other hand are noisy.  In those architectures where the single-qubit gates contribute a significant error (see e.g.~\cite{morvan2023phase}), it can be treated in the same manner as we treat the two-qubit gate errors below.

We consider a simplistic noise model where the application of any two-qubit gate is followed by a transformation:

\begin{equation}
    \label{eq:channel_type}
    \rho \rightarrow (1 - q) \rho + q \mathcal{E}(\rho),
\end{equation}
where $q$ is the gate error rate and $\mathcal{E}$ is a completely positive trace-preserving (CPTP) map applied to the pair of qubits on which the gate operates. 

We consider inhomogeneous errors associated with two-qubit gates.  We denote the gate error rate associated with $G^l_{jk}$ to be $q^l_{jk}$. We further assume that the error rate for a gate only depends on the qubit pair it acts on, that is $q^{l}_{jk} = q_{jk}, ~\forall (j,k)$. For simplicity our analysis neglects the role of state preparation and measurement (SPAM) errors and cross-talk errors.
Thus, the set $\{q_{jk}\}$ and the channel $\mathcal{E}$ describe the error model completely. 

Recall that the action of the map in \eqref{eq:channel_type} on a given state is tantamount to applying $\mathcal{E}$ with probability $q$, and operating trivially with probability $(1-q)$. Following this argument one can show that if $\rho_0$ is the state that one expects to be prepared by the noiseless circuit, one would instead obtain the state:
\begin{equation}
\label{eq:npnst}
    \rho = \sum_{s} \left( \prod_{\substack{(j,k) \in T\\ l \in [1,d]}} (1 - q_{jk})^{1 - s^l_{jk}} (q_{jk})^{s^l_{jk}} \right)    \rho_s.
\end{equation}
Here $s$ is a two dimensional array with elements $s^l_{jk} \in \{0,1\}$, that indexes the two-qubit gates after which $\mathcal{E}$ is appended: if $s^l_{jk} = 1$ the channel $\mathcal{E}$ is appended after the application of $G^l_{jk}$ while if $s^l_{jk} = 0$ the state remains intact after the application of $G^l_{jk}$. Therefore $\rho_s$ is the state obtained from $\rho_0$ by appending the error channels $\mathcal{E}$ as determined by $s$; $\rho_0 = \rho_{s}$ with $s^l_{jk} = 0~\forall (j,k) \in T, l \in [1,d]$.

\subsection{Perturbative analysis}
\label{sec:perturb_theory}

In this section we consider $q_{jk}$ to be small perturbative terms, such that $(\max\{q_{jk}\}) \vert T \vert d \ll 1$. Here $\vert \cdot \vert$ represents the cardinality of a set. This allows us to discard all terms that are at least quadratic in $q_{jk}$. The Taylor expansion of \eqref{eq:npnst} in $q_{jk}$ yields the linear approximation to $\rho$:
\begin{equation}
    \rho = \left(1 - d\sum_{(j,k) \in T} q_{jk} \right)\rho_{0} + \sum_{\substack{(j,k) \in T \\ l \in [1,d]}} q_{jk} \rho^l_{jk} + O(q^2).
\end{equation}
where $\rho^l_{jk} = \rho_{s}$ such that $s$ has only one non-zero entry $s^l_{jk} = 1$.

Let us now consider the behavior of the energy $E = \Tr \rho H$ with respect to the error rates: 
\begin{multline}
        \label{eq:e_vs_q_linear_avg}
    E - E_0 = -d E_0 \sum_{(j,k) \in T} q_{jk} + \sum_{\substack{(j,k) \in T \\ l \in [1,d]}} q_{jk} E^l_{jk} + O(q^2) \\
    = d (A - E_0)  \sum_{(j,k) \in T} q_{jk}  + \sum_{\substack{(j,k) \in T \\ l \in [1,d]}} q_{jk} \epsilon^l_{jk} + O(q^2).
\end{multline}
Here, $E_0 = \Tr \rho_{0} H$ is the energy of the ansatz state in the absence of noise, 
$E_{jk}^l = \Tr \rho^l_{jk} H$ are the energies obtained by applying an error channel $\mathcal{E}$ to qubits $(j, k)$ in the $l$-th layer of the ansatz,
$A= \frac{1}{\vert T \vert d} \sum E_{jk}^l$, and $\epsilon^l_{jk} = E_{jk}^l - A$.

Equation~(\ref{eq:e_vs_q_linear_avg}) has two $q_{jk}$-dependent terms. The first term shows a linear dependence of $(E-E_0)$ with respect to the circuit error sum $d \sum q_{jk}$. The second term, on the other hand, quantifies the deviation from the linear trend. In Appendix \ref{appen_0} we establish bounds on the relative deviation and show that under specific conditions $(E-E_0)$ behaves linearly with respect to the circuit error sum. 

\subsection{ZNE with permutation fit}
\label{sec:explain_fit}

In this section we will lay down the recipe for ZNE assisted by inhomogeneous errors. Consider once again the variational circuit which we now want to implement on a physical device. Here one must make a distinction between the abstract qubits (ones in the circuit) and the physical qubits (ones in the device). 

We begin by assuming that it is possible to implement two-qubit gates on all pairs of physical qubits and we denote the corresponding error rates as $\bar{q}_{jk};j,k \in [1,n]$. 
Later in the paper we also consider cases of limited connectivity between physical qubits. 

\begin{figure}[htp]
    \centering
    \includegraphics[width=\linewidth]{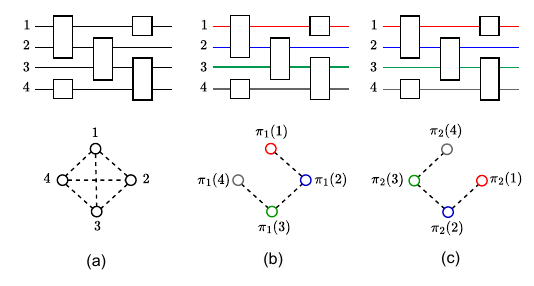}
    \caption{(color online) (a) A 4-qubit (abstract) circuit (top) to be executed on a 4-qubit (physical) device (bottom) with all-to-all connectivity, depicted with dashed lines. (b) and (c) two different abstract-to-physical qubit mappings $\pi_1$ and $\pi_2$ (also indicated by colors). The corresponding qubit connections used for circuit implementation are indicated with dashed lines. Each of these mappings corresponds to a different circuit error sum.}
    \label{fig:mapping}
\end{figure}

In order to implement an $n$-qubit (abstract) circuit on an $n$-qubit (physical) device, one would first require to map the abstract qubits to their physical counterparts---what we call abstract-to-physical qubit mapping. Mathematically the mapping can be specified by a permutation $\pi \in S_n$ such that an abstract qubit indexed $j$ is mapped to the physical qubit indexed $\pi(j)$. Examples of different mappings are demonstrated in Fig. \ref{fig:mapping}. Under a chosen qubit mapping the error rate associated with the gate $G^l_{jk}$, operating on the abstract qubit pair $(j,k)$ and hence the physical qubit pair $(\pi(j), \pi(k))$, will be denoted as $\bar{q}_{\pi(j) \pi(k)} \equiv \bar{q}_{\pi(jk)}$. 
For this choice of qubit mapping, the energy of the VQE ansatz state (\ref{eq:e_vs_q_linear_avg}) transforms into the following:

\begin{equation}
    \label{eq:e_permute}
    E - E_0 =  d (A - E_0) \sum_{(j,k) \in T} \bar{q}_{\pi(jk)}  + \sum_{\substack{(j,k) \in T \\ l \in [1,d]}} \bar{q}_{\pi(jk)} \epsilon^l_{jk} + O(q^2).
\end{equation}

Evidently, for a generic ansatz circuit the sum $\sum \bar{q}_{\pi(jk)}$ will depend on the permutation $\pi$ and hence on the abstract-to-physical qubit mapping. Thus, by taking a number of permutations, we obtain the energies for different circuit error sums which can be approximated by a linear dependence of the form $E = a \sum \bar{q}_{\pi(jk)} + b$, where the slope $a$ approximates the term $d (A - E_0)$, and the interception point $b$ approximates the noise-free energy $E_0$. Consequently, by making a linear fit through the noisy data, one could recover an approximation to the noiseless energy $E_0$. In general one cannot guarantee that the estimate constructed this way is exact. However, in Appendix \ref{intercept_bnd} we show that this method of ZNE, when performed over all permutations $\pi \in S_n$, can recover the exact value of the noise-free energy for ansatz structures corresponding to regular graphs, in the limit of small errors $\bar{q}_{jk}$. 

\section{Numerical results}
\label{sec:numerics}

In this section we demonstrate the performance of ZNE applied to noise-perturbed VQE circuits. To investigate the applicability range of the protocol, we use it for a number of different problem Hamiltonians, noise models and error distributions.

We consider three types of Hamiltonians. The first two are variants of the transverse field Ising model:
\begin{equation}
\label{eq:tfim}
    H = \sum_{j=1}^n J_j Z_j Z_{j+1} + \sum_{j=1}^n h_j X_j,
\end{equation}
where $Z_{n+1} \equiv Z_{1}$. Two cases are given by (a) $J_j = h_j = 1$ and (b) $J_{1} = 6$, $J_j = 1~\forall j \neq 1$ and $h_j = 1$. We will refer to these as Ising A and Ising B. 

The third type of Hamiltonian is constructed from the electronic structure model of the $\mathrm{H}_2\mathrm{O}$ molecule. We construct the second quantized Hamiltonian in the ccpvdz basis using the pyscfdriver module of Qiskit, then choose different sizes of the active space to create problems of different sizes. Finally, we apply the Bravyi-Kitaev transformation and apply a penalty term to ensure the correct particle number:
\begin{equation}
    H_{p} = H + \mu (\hat{n} - n_e)^2.
\end{equation}
Here $\hat{n}$ is the particle number operator $\sum a_i^\dagger a_i$ transformed with the Bravyi-Kitaev transformation, and $n_e$ is the desired number of electrons. We choose the weight of the penalty term $\mu$ according to the ``rough'' rule from ~\cite{kuroiwa_penalty_2021-1}, which means that $\mu$ is equal to twice the sum of absolute values of the weights of Pauli strings in $H$. This choice ensures that the ground state of $H_p$ has the correct particle number.

We minimize the energy of the Hamiltonians with respect to variational states prepared by a Hardware Efficient Ansatz (HEA) with ring and line topology. For both topologies, in every layer of the ansatz we apply single-qubit gates to each qubit ($R_Y$, then $R_X$) and then apply two-qubit $R_{ZZ}$ gates to nearest neighboring qubits. For the ring topology we apply an additional $R_{ZZ}$ gate to the qubit pair $(n, 1)$. In the rest of the paper we mostly focus on the results for the ring topology, as both ansatze gave qualitatively similar results. 

The optimization is done in the Qiskit statevector simulator using the Limited memory Broyden–Fletcher–Goldfarb–Shanno (L-BFGS-B) algorithm. The starting point for the optimization is chosen by picking every parameter at random from the normal distribution with mean zero and $\sigma=10^{-3}$. To improve convergence, we use a variant of layerwise learning~\cite{zhou_quantum_2020-1,skolik_layerwise_2021}: first we start with a one-layered ansatz and optimize it. Then we add another layer, initialized at parameter values close to zero, and optimize the new ansatz, then continue the same procedure with the addition of each subsequent layer. Before every new layer we slightly perturb the solution to avoid the saturation effect~\cite{Campos2021}.

Using an approximately optimal ansatz state, we simulate the ZNE protocol. For that, each two-qubit gate $g_{jk}$ is appended with a local noise channel with random strength $\Bar{q}_{jk} = \Bar{q}_{kj}$. A number of random qubit permutations is sampled, and for each permutation the CES and the noisy VQE energy are calculated. Finally, this data is used to make a linear fit and obtain the error-mitigated value of the energy.

\subsection{Zero-noise extrapolation}
\label{sec:zne_general}

We begin by considering a case of $n=6$ qubits with  depolarizing noise of strength $\{\bar{q}_{jk}\}$ sampled from a uniform distribution on the interval $[0, 0.001]$. The depth of the ansatz circuit is chosen such that the noiseless VQE energy differs from the ground energy of the problem by no more than $1\%$ of the spectral gap. In the considered case, the required depths were 4, 8 and 1 layers for the Ising A, Ising B, and water Hamiltonians, respectively.

The results of ZNE for all considered Hamiltonians are demonstrated in Fig.~\ref{fig:epsilons_and_fits}. 
Each point in Fig.~\ref{fig:epsilons_and_fits} (a) - \ref{fig:epsilons_and_fits} (c) represents energy $E$ obtained for a specific permutation $\pi \in S_n$; all possible 720 permutations were used for ZNE. A linear fit of energies for the considered perturbations indeed allows one to recover the energy of the noiseless VQE approximation with a high accuracy. In all three cases, the extrapolated energy differs from the true ground energy by at most $5\times10^{-4}$. Indeed, in Appendix~\ref{intercept_bnd} we show that, in the case of ansatz circuits with a regular graph structure, the ZNE over all permutations approximates noiseless expectation value exactly (up to higher order terms in \eqref{eq:e_permute}).


The energies obtained for the Ising A model and the water Hamiltonian better approximate a linear trend compared to the energies obtained for the Ising B model. To explain this behavior we recall that the deviation from the linear trend is governed by the non-uniformity of the energies $\epsilon_{jk}^l$, which is captured by the quantity $\text{max}|\sum_{l} \epsilon^l_{jk} |$ (see Appendix \ref{appen_0}). Figs.~\ref{fig:epsilons_and_fits} (d) - (f) illustrate that point: for the Ising A model and water Hamiltonians the energies $E_{jk}^l$ are more concentrated around the mean, and the linear trend with the CES is more apparent. At the same time, for the Ising B model the distribution of the energies is wider, and consequently, the linear trend is less evident.

\begin{figure}[htp]
    \centering
    \includegraphics[width=\linewidth]{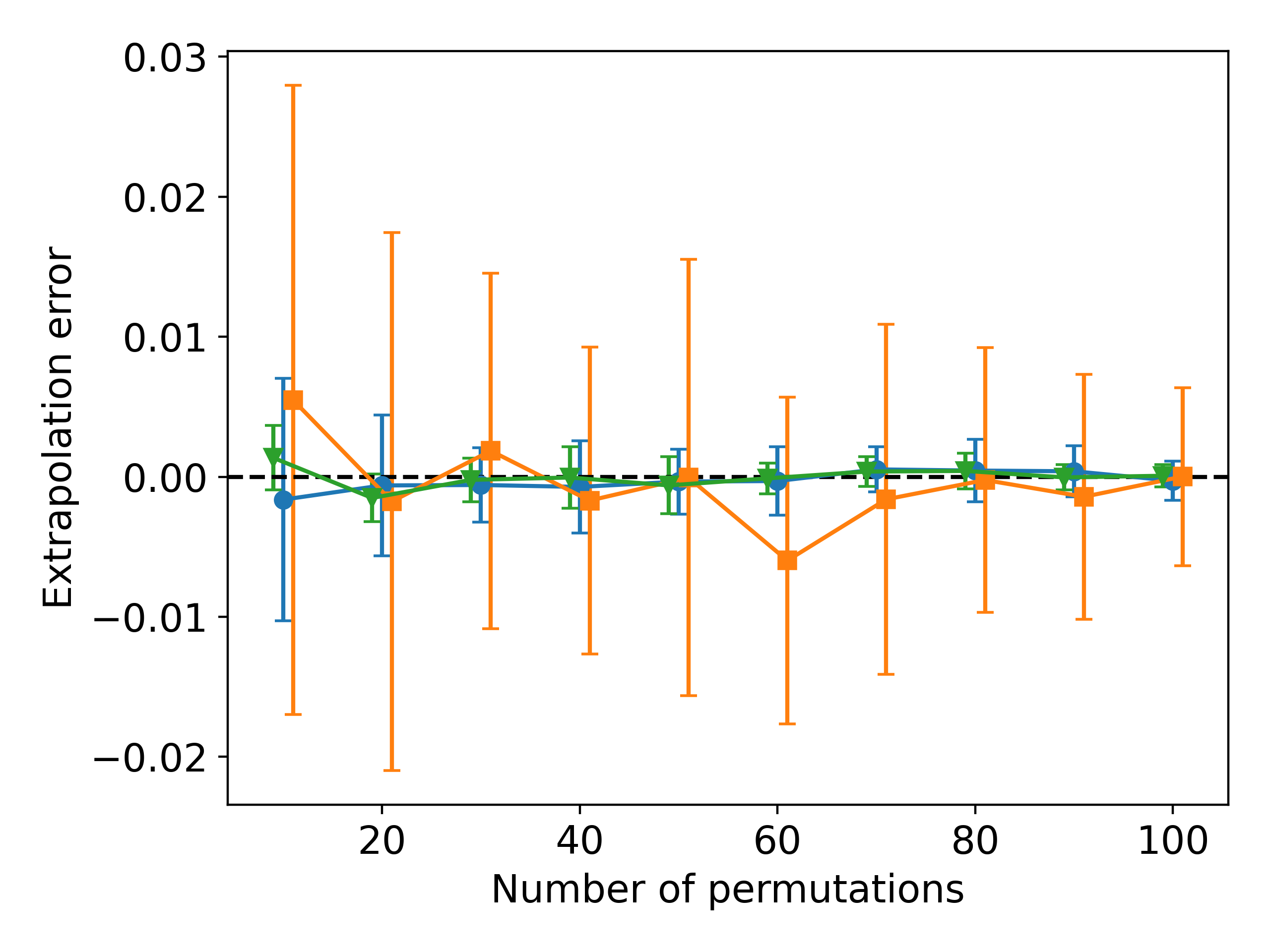}
    \caption{ZNE error versus the number of permutations used. Blue circles: Ising A, orange squares: Ising B, green triangles: $\mathrm{H}_2\mathrm{O}$. The markers for Ising B and water are offset horizontally for visibility. The error bars indicate standard deviation from considering 20 different noise realizations (sets of $\{\bar{q}_{jk}\}$ sampled). For every realization a different set of randomly generated permutations was considered.}
    \label{fig:stub_1}
\end{figure}

Deviation from the linear trend, however, does not necessarily worsen the quality of the approximation, when all permutations are considered. Indeed, notwithstanding any deviation, the ring topology allows to recover an exact energy approximation (see Appendix \ref{intercept_bnd}).
However, taking all permutations is not feasible except for the smallest problems. In addition, in many quantum computer architectures, implementing an arbitrary permutation of a circuit is impossible without introducing additional swap gates. For example, while ion-based quantum processors have all-to-all qubit connectivity, quantum processors based on superconducting qubits or trapped atoms do not.
To address that, we investigated the performance of ZNE when only a limited number of permutations is available.  Fig.~\ref{fig:stub_1} shows the performance of ZNE as a function of the size of the permutation pool. 
For each of the three Hamiltonians, the uncertainty in the ZNE energy increases for smaller size of the permutation pool. However, the energies are still centered around the noise-free energy, with the standard deviation being much smaller than any energy scale of the problem in each case.

\begin{widetext}

\begin{figure}[h]
\begin{minipage}[h]{0.31\linewidth}
\center{\includegraphics[width=1\linewidth]{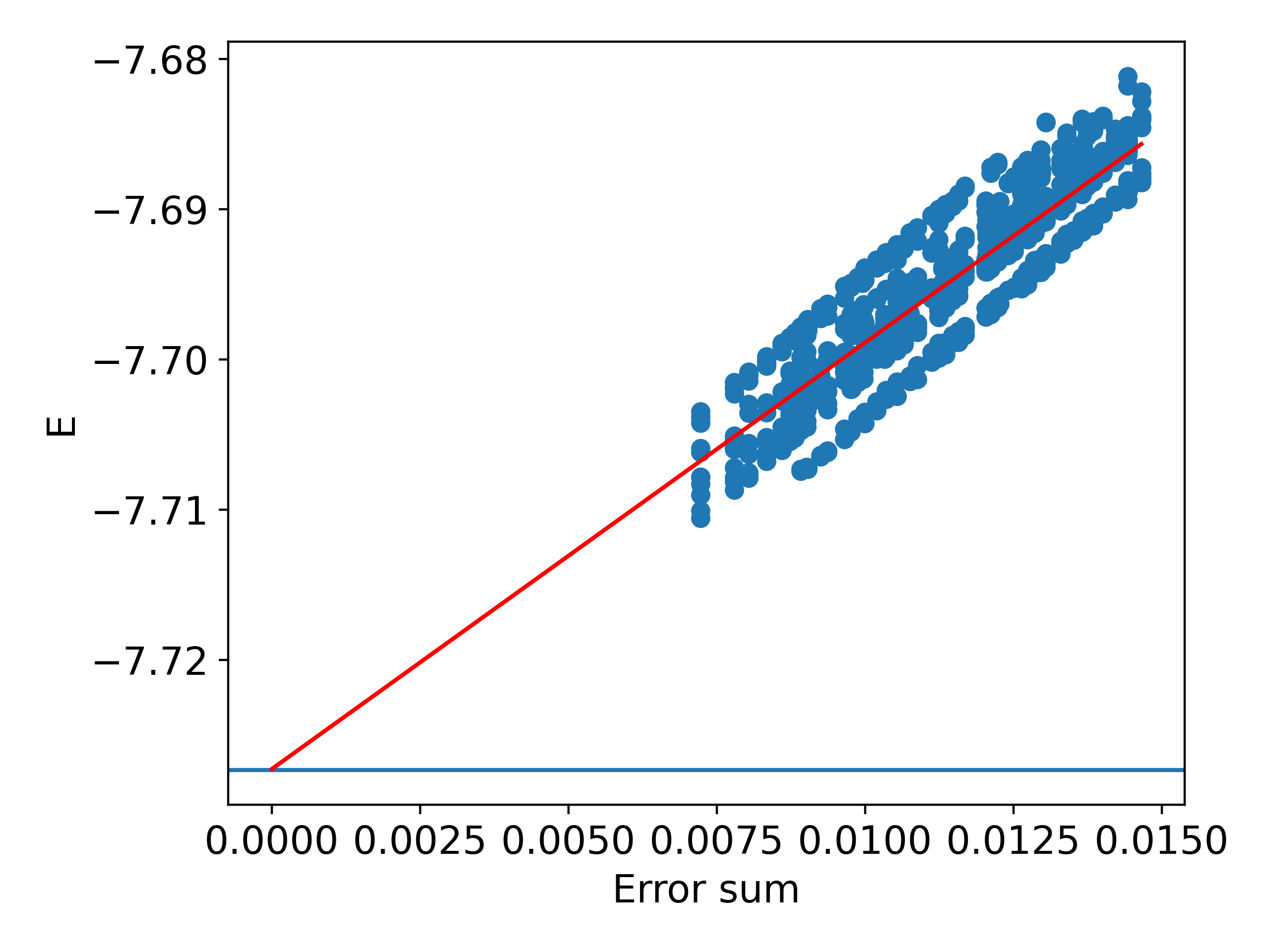}} (a) \\
\end{minipage}
\hfill
\begin{minipage}[h]{0.31\linewidth}
\center{\includegraphics[width=1\linewidth]{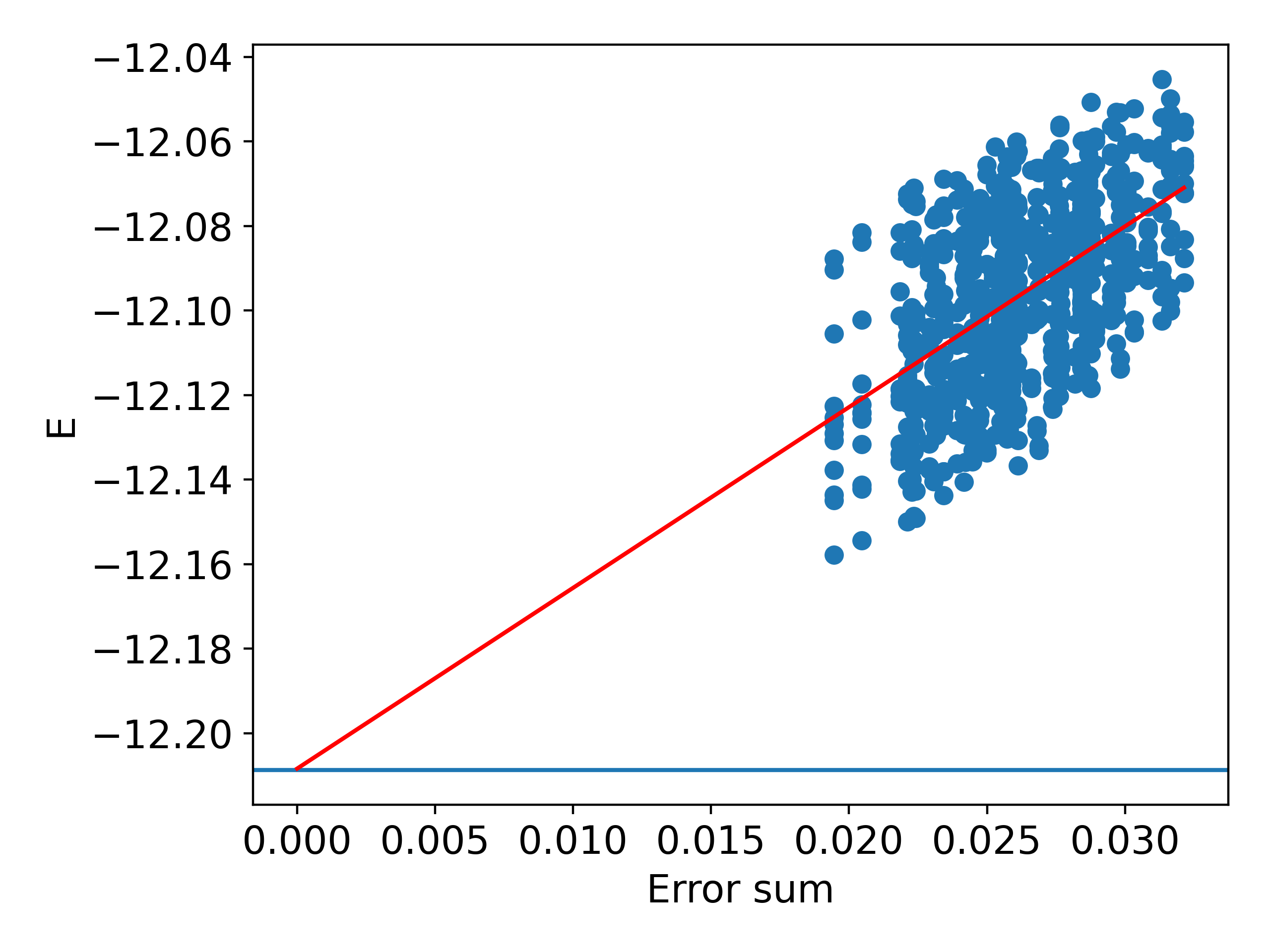}} \\(b)
\end{minipage}
\hfill
\begin{minipage}[h]{0.31\linewidth}
\center{\includegraphics[width=1\linewidth]{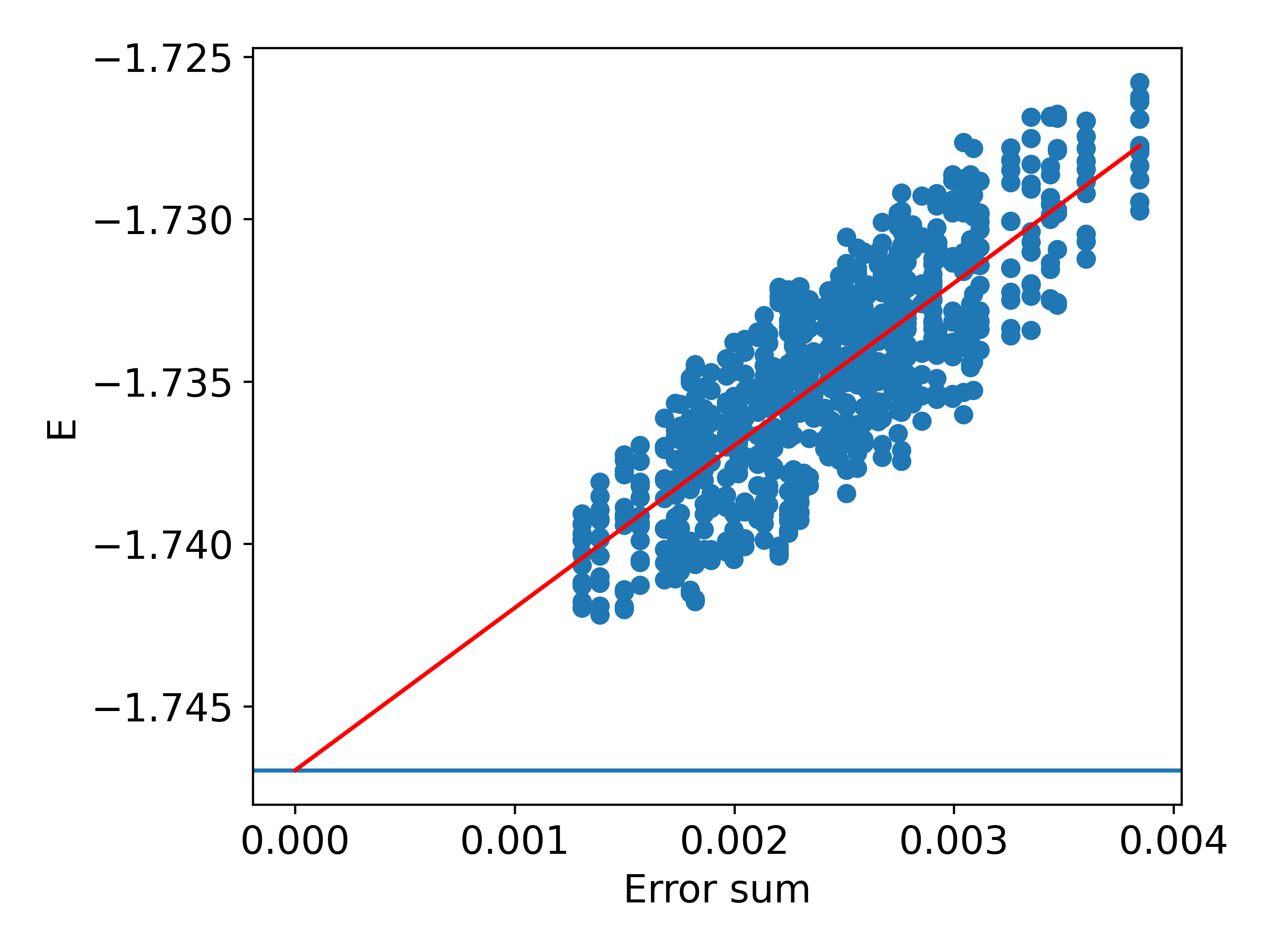}} \\(c)
\end{minipage}
\vfill
\begin{minipage}[h]{0.31\linewidth}
\center{\includegraphics[width=1\linewidth]{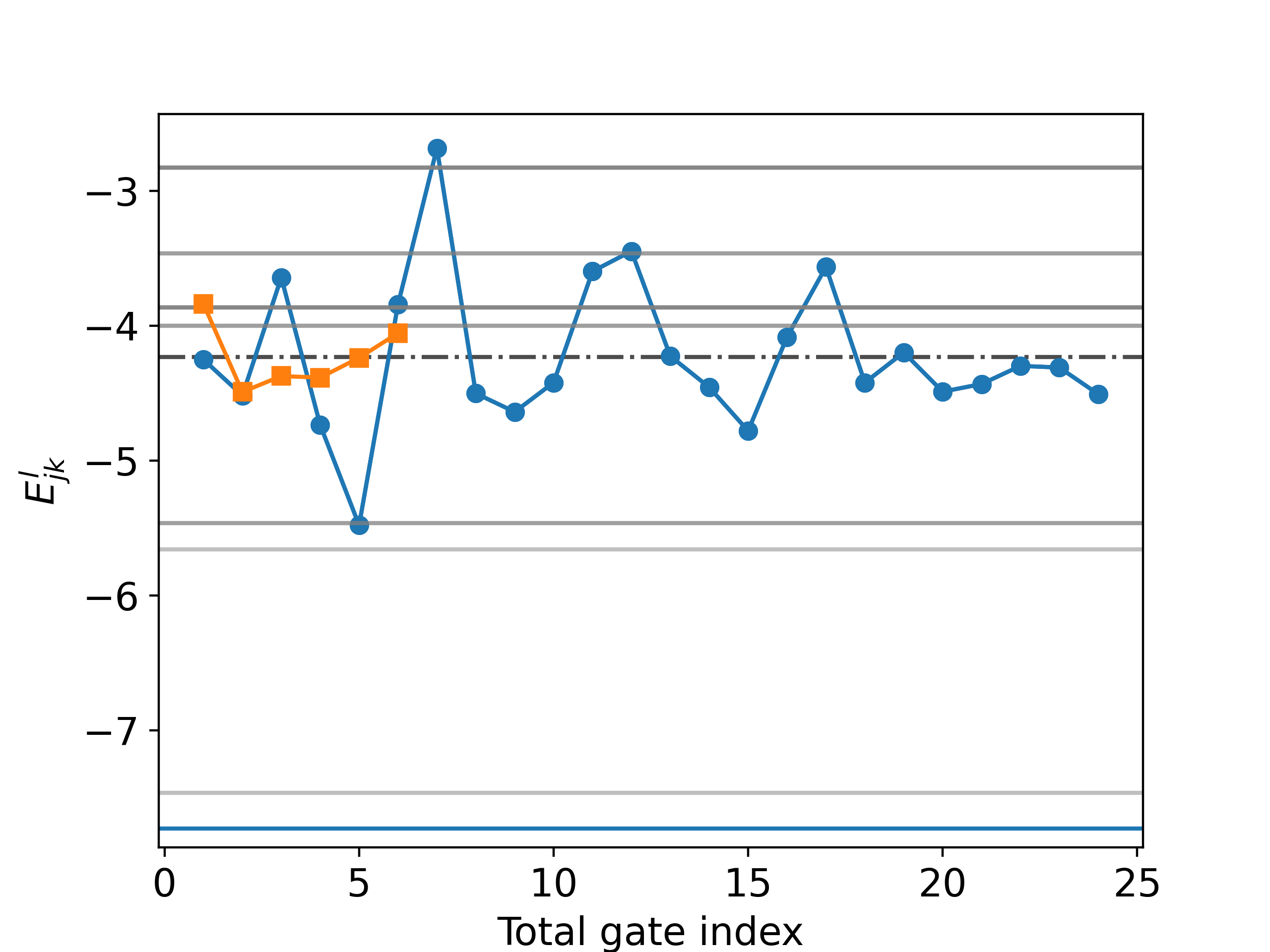}} \\(d)
\end{minipage}
\hfill
\begin{minipage}[h]{0.31\linewidth}
\center{\includegraphics[width=1\linewidth]{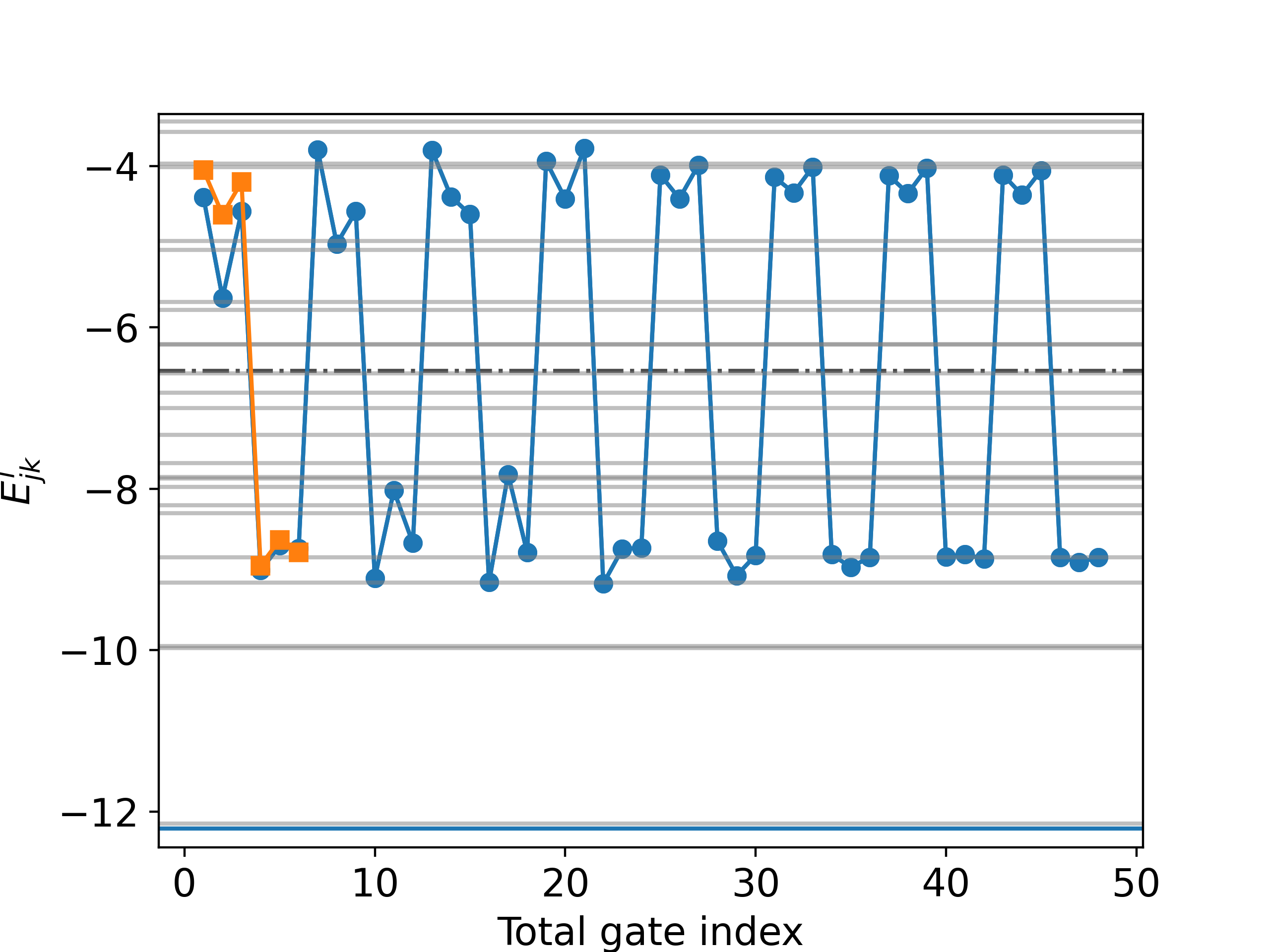}} \\(e)
\end{minipage}
\hfill
\begin{minipage}[h]{0.31\linewidth}
\center{\includegraphics[width=1\linewidth]{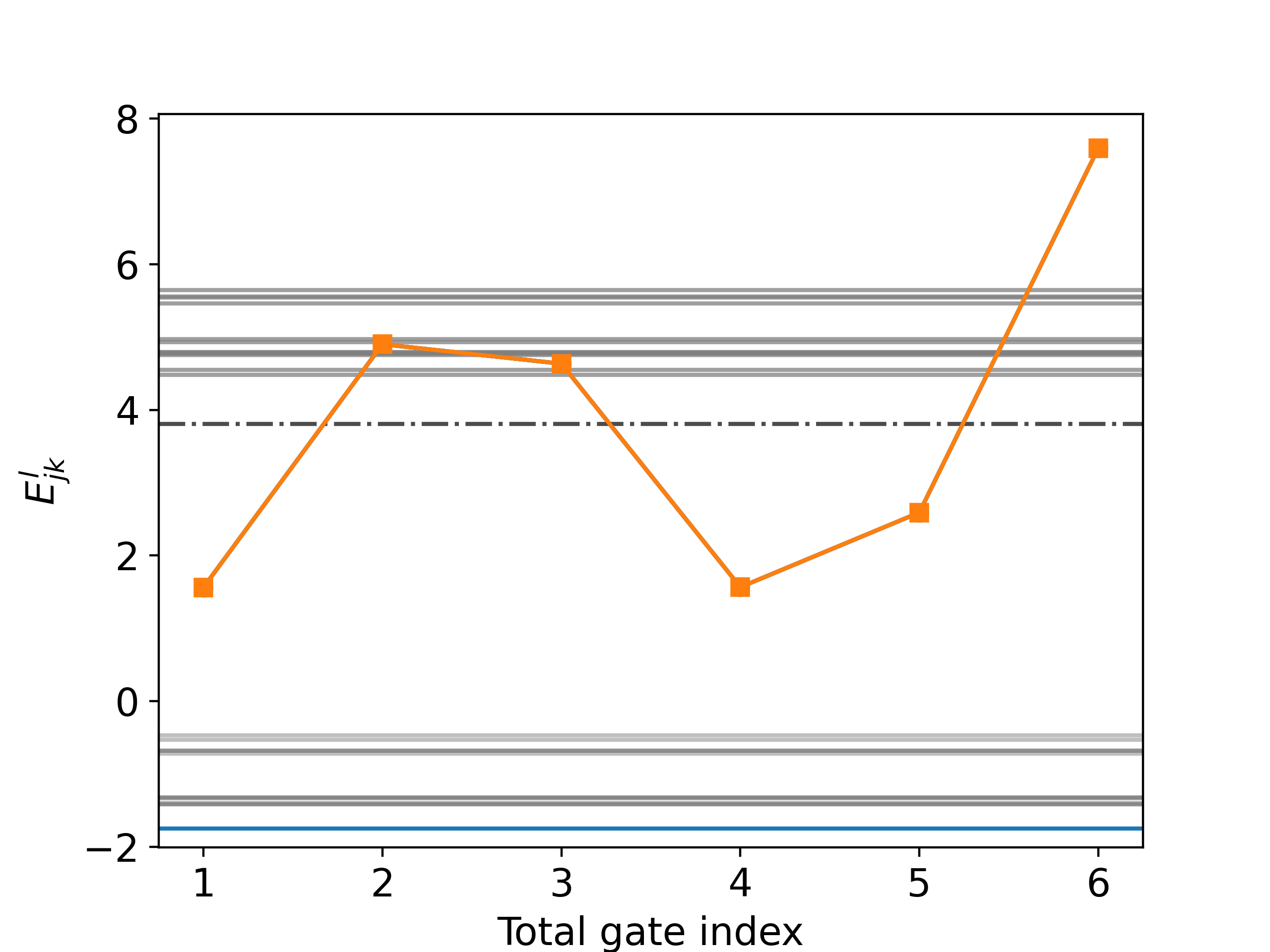}} \\(f)
\end{minipage}
\caption{(a), (b), (c): Zero-noise extrapolation (ZNE) using the proposed method
for $n=6$ qubit Ising A, Ising B, and water CAS Hamiltomians, respectively. Blue blue dots on these plots represent energy $E$ as per \eqref{eq:e_permute} for different permutations $\pi \in S_n$ and the red line is a linear fit taken over energies corresponding to all possible permutations. (d), (e), (f): ~The energies $E_{jk}^l$ (see \eqref{eq:e_vs_q_linear_avg}) for Ising A, Ising B, and water Hamiltonians respectively (shown in circles). The squares show the perturbation energies averaged over the layers of the ansatz $\frac{1}{d}\sum_l E_{jk}^l$. The gray lines represent the eigenvalues of the Hamiltonian. The dash-dotted lines depict the average perturbation energy $A$.
On all the panels blue horizontal lines show the noiseless VQE energy.}
\label{fig:epsilons_and_fits}\end{figure}
\vspace{2em}
\end{widetext}

Notwithstanding the success of the protocol with limited number of permutations, we investigate a specific case where available permutations are dictated by connectivity of the hardware.
For this we ran an experiment with $n=6$ qubits assuming that the connectivity graph is a $2 \times 3$ square lattice. In this case, we consider an ansatz of the line topology and embedded it in all 16 possible ways into the device. We ran ZNE over 20 random realizations of the noise for the Ising A problem and found that the mean extrapolation error is $-3.6 \times 10^{-4}$ units with standard deviation $2.1 \times 10^{-3}$. This demonstrates that the proposed ZNE protocol can work adequately withing the practical limitations of hardware connectivity.

\subsection{Scaling with noise strength and problem size}
\begin{figure}
    \includegraphics[width=\linewidth]{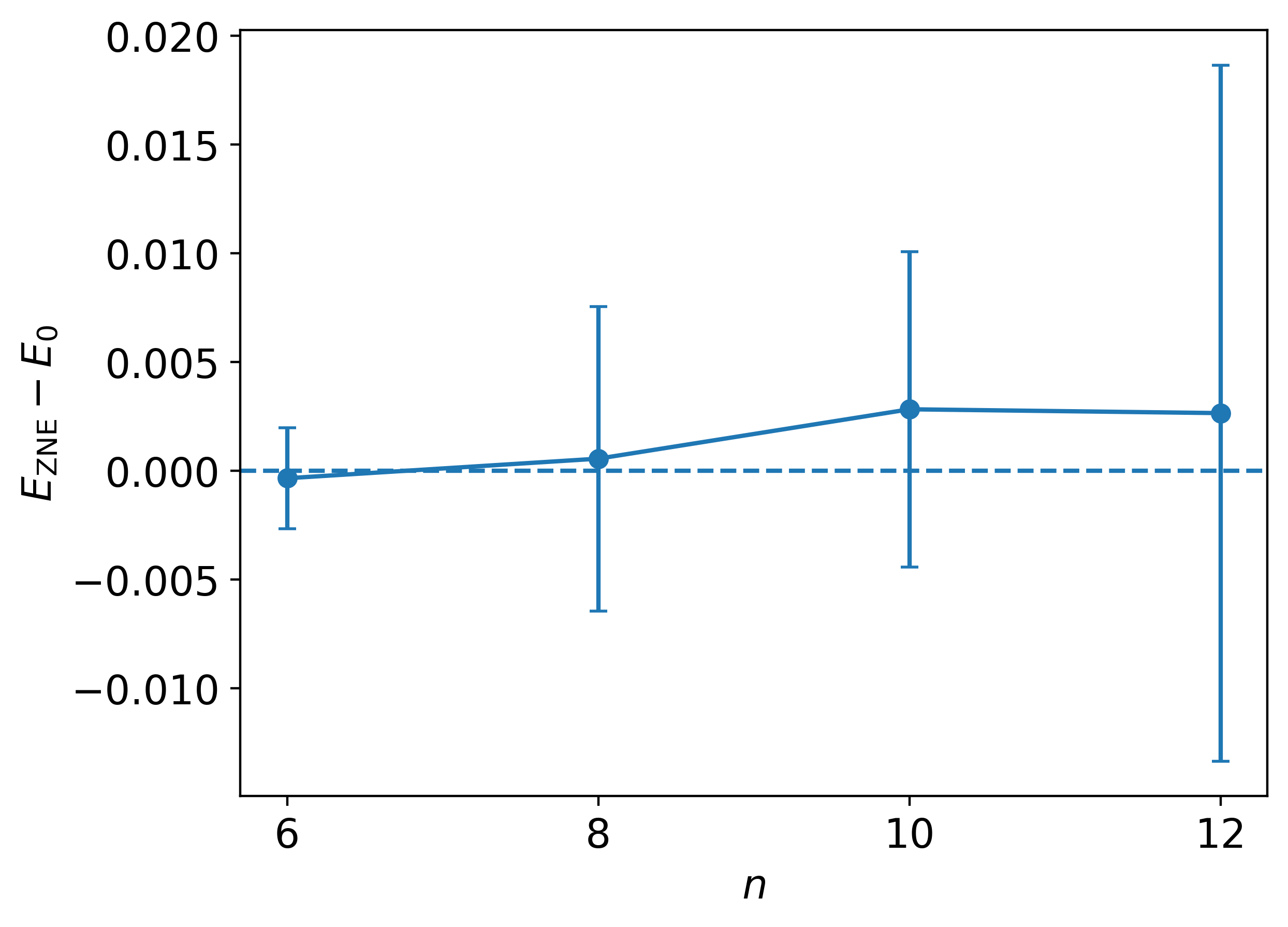}
    \caption{Scaling of the extrapolation error for depolarizing noise with $\langle q\rangle = 5\times10^{-4}$ with respect to $n$. The mean values and deviations are estimated from considering 20 different noise realizations.}
    \label{fig:zne_error_n}
\end{figure}

The proposed method relies on the inhomogeneity of the error rates, essentially taking sums of error rates over particular subsets of qubit pairs. For larger system sizes and circuits, however, the sum of error rates converges towards the mean, leading to the concentration of circuit error sums. This might induce an instability in the ZNE estimated energy.
To investigate this, we perform ZNE for system sizes $n \in \{6, 8, 10, 12\}$ for Ising A Hamiltonian, using 50 random permutations and averaging the results over 20 different noise realizations of average strength $\langle q \rangle = 5\times 10^{-4}$ (corresponding to uniform sampling from $[0, 0.001]$). The depth of the noiseless VQE circuit again is taken to approximate the ground energy within $1\%$ of the spectral gap. The resulting distributions of ZNE extrapolation error are shown in Fig.~\ref{fig:zne_error_n}. We note that the standard deviation increases gradually with respect to $n$. One can offset this increase in the standard deviation by averaging the ZNE energy over at most polynomially (with respect to $n$) many experiments.

\begin{figure}
        \includegraphics[width=\linewidth]{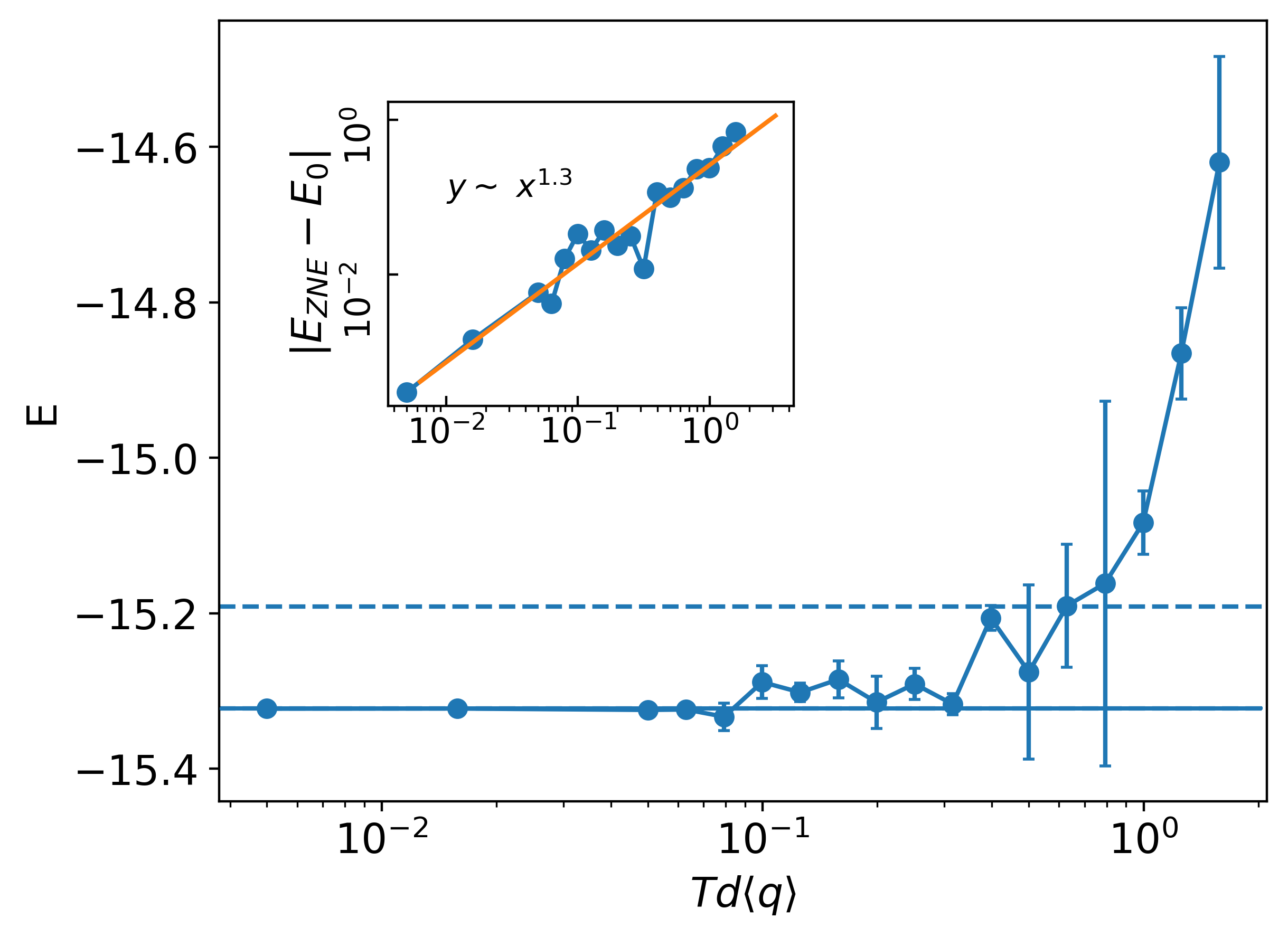}

  \caption{ZNE estimated energy as a function of the magnitude of circuit errors for $n=12$ qubits. The mean values and deviations are estimated from considering 3 different noise realizations. Solid horizontal line shows the noise-free VQE energy, dashed line is separated from it by the value of the spectral gap.}
  \label{fig:zne_error_a}
  
\end{figure}

In the theoretical analysis, we demanded that $(\max\{q_{jk}\}) \vert T \vert d \ll 1$ in order to discard the quadratic terms. However, this requirement can be restrictive for practical purposes. Thus, we  investigate the real applicability range of the method. The behavior of the ZNE approximation for $n = 12$ qubit Ising A Hamiltonian across a wide range of mean noise strength $\langle q \rangle$ is shown in Fig.~\ref{fig:zne_error_a}. It is seen that the ZNE error is smaller than the spectral gap of the problem up to noise threshold value $\vert T \vert d \langle q\rangle_{thr}\sim 1$. Moreover, even for larger values of $\vert T \vert d\langle q\rangle$, the extrapolation was closer to the true ground state energy than any of the noisy energy values. 
In general, the absolute value of the extrapolation error appears to scale polynomially with the amplitude of the noise (Fig.~\ref{fig:zne_error_a}, inset).

\subsection{ZNE with different noise models}
The success of the proposed ZNE protocol may depend on the distribution of error rates, as well as on the considered noise model. To investigate this, we simulate ZNE for (i) two-qubit depolarizing noise, as well as for (ii) more realistic two-qubit Pauli error \cite{Wallman2016}. For each noise model the error rates are sampled from (a) uniform and (b) more realistic lognormal distributions \cite{Kim2023, morvan2023phase}. 
To make comparison possible, the distributions were normalized to have the same means and variances. For the Pauli error, the means were divided by 15, so that the sum of the error coefficients has the same mean and variance. 

For each noise type we simulate the experiments described in the previous subsection. For the noiseless VQE we considered circuits of up to 10 layers. As before, for most of the problems this was enough to find energy approximation within $ 1\%$ of the spectral gap from the true ground state. The exceptions were  Ising B for $n=10$ and $n=12$, where the error was about 0.5 of the gap, and $\mathrm{H_2 O}$ for $n=12$, where the error was approximately equal to the gap. 

In Table \ref{tab:thresholds} we present results of ZNE using 50 random permutations and averaged over 20 noise realizations for the discussed noise types and system sizes. Here $\Delta E$ represents ZNE error for a mean noise strength $\langle q\rangle = 5\times 10^{-4}$. It can be seen that ZNE for different noise models and error distributions demonstrate qualitatively similar results, thus indicating the potential of the proposed protocol. The error growth for larger system sizes is caused by the fixed number of permutations considered, and can be countered by its increase, in accordance with Fig \ref{fig:stub_1}. 

In addition, in Table \ref{tab:thresholds} we summarize the threshold strength of the noise $\langle q\rangle_{thr}$, below which ZNE estimated energy differs from noiseless VQE energy by no more than the spectral gap. Those results were obtained by simulating ZNE with 50 permutations across a wide range of noise amplitudes,  averaging the results over 3 noise realizations, in the same manner as depicted in Fig \ref{fig:zne_error_a}. By interpolating the resulting curves we identify the threshold values $\langle q\rangle_{thr}$, at which the curve crosses spectral gap. 
Again, it is seen that noise strength $ \vert T \vert d \langle q\rangle_{thr}\lesssim 1$ allows one to recover ZNE energy withing one spectral gap from the noiseless VQE. Moreover, even for larger values of $ \vert T \vert d\langle q\rangle$ the extrapolation is closer to the true ground state energy than any of the noisy energy values.

\textbf{\begin{table*}[ht!]
    \caption{\label{tab:thresholds} Experimental results for ZNE using 50 permutations with different Hamiltonians and noise models. $\Delta E$ gives ZNE error produced for mean noise strength $\langle q\rangle = 5\times 10^{-4}$ (corresponding to sampling from $[0, 0.001]$ in the case of uniform distribution). The results are averaged over 20 noise realizations. $q_{thr}$ represents the mean noise strength at which ZNE estimation exceeds noiseless VQE energy by the value of spectral gap. 
    The results are averaged over 3 noise realizations.}
    \centering
    \begin{tabular}{|c|c|c|c|c|c|c|}
    \hline
    Noise type & \multicolumn{6}{|c|}{Hamiltonian} \\
    \cline{2-7}  
    & \multicolumn{2}{|c|}{Ising A} 
    & \multicolumn{2}{|c|}{Ising B} 
    & \multicolumn{2}{|c|}{$\mathrm{H_2 O}$} \\
    \cline{2-7}
    & $\Delta E$ & $|T|d\langle q\rangle_{thr}$
    & $\Delta E$ & $|T|d\langle q\rangle_{thr}$
    & $\Delta E$ & $|T|d\langle q\rangle_{thr}$ \\    
    \hline
    \multicolumn{7}{|c|}{$n=6$} \\
    \hline
    Depolarizing, uniform & $(-3.4 \pm 23.2) \cdot 10^{-4}$ & $0.83 \pm 0.11$ & $(0.0 \pm 1.6) \cdot 10^{-2}$ & $0.14 \pm 0.04$ & $(-0.6 \pm 2.0) \cdot 10^{-3}$ & $0.77 \pm 0.15$\\
    Depolarizing, lognormal & $(-1.4 \pm 26.3) \cdot 10^{-4}$ & $0.35 \pm 0.03$ & $(-0.2 \pm 1.0) \cdot 10^{-2}$ & $0.09 \pm 0.03$ & $(0.2 \pm 1.2) \cdot 10^{-3}$ & $0.50 \pm 0.17$\\
    Pauli, uniform & $(2.2 \pm 5.2) \cdot 10^{-3}$ & $0.65 \pm 0.10$ & $(0.2 \pm 2.0) \cdot 10^{-2}$ & $0.10 \pm 0.03$ & $(0.1 \pm 5.5) \cdot 10^{-3}$ & $0.41 \pm 0.14$\\
    Pauli, lognormal & $(-7.3 \pm 43.9) \cdot 10^{-4}$ & $0.33 \pm 0.04$ & $(0.0 \pm 2.1) \cdot 10^{-2}$ & $0.06 \pm 0.02$ & $(0.4 \pm 5.2) \cdot 10^{-3}$ & $0.15 \pm 0.03$\\     
    \hline
    \multicolumn{7}{|c|}{$n=8$} \\
    \hline
    Depolarizing, uniform & $(0.5 \pm 7.0) \cdot 10^{-3}$ & $0.81 \pm 0.11$ & $(0.2 \pm 3.6) \cdot 10^{-2}$ & $0.18 \pm 0.05$ & $(0.6 \pm 2.2) \cdot 10^{-2}$ & $0.18 \pm 0.02$\\
    Depolarizing, lognormal & $(-0.8 \pm 4.2) \cdot 10^{-3}$ & $0.33 \pm 0.03$ & $(-0.4 \pm 3.6) \cdot 10^{-2}$ & $0.09 \pm 0.01$& $(-0.2 \pm 2.1) \cdot 10^{-2}$ & $0.13 \pm 0.01$\\
    Pauli, uniform & $(0.2 \pm 7.8) \cdot 10^{-3}$ & $0.53 \pm 0.04$ & $(-0.8 \pm 5.9) \cdot 10^{-2}$ & $0.13 \pm 0.03$ & $(0.7 \pm 3.6) \cdot 10^{-2}$ & $0.07 \pm 0.01$\\
    Pauli, lognormal & $(0.2 \pm 1.0) \cdot 10^{-2}$ & $0.31 \pm 0.02$ & $(-0.1 \pm 5.5) \cdot 10^{-2}$ & $0.06 \pm 0.01$& $(0.2 \pm 5.2) \cdot 10^{-2}$ & $0.07 \pm 0.02$\\  
    \hline
    \multicolumn{7}{|c|}{$n=10$} \\ \hline
    Depolarizing, uniform & $(2.8 \pm 7.2) \cdot 10^{-3}$ & $0.72 \pm 0.06$ & $(-0.6 \pm 1.7) \cdot 10^{-2}$ & $0.14 \pm 0.04$& $(0.6 \pm 4.7) \cdot 10^{-2}$ & $0.10 \pm 0.03$\\
    Depolarizing, lognormal & $(0.0 \pm 1.0) \cdot 10^{-2}$ & $0.30 \pm 0.01$ & $(-0.3 \pm 1.5) \cdot 10^{-2}$ & $0.09 \pm 0.02$& $(-1.3 \pm 4.7) \cdot 10^{-2}$ & $0.09 \pm 0.02$\\
    Pauli, uniform & $(0.5 \pm 1.5) \cdot 10^{-2}$ & $0.51 \pm 0.04$ & $(-0.3 \pm 2.8) \cdot 10^{-2}$ & $0.08 \pm 0.02$& $(-1.6 \pm 8.6) \cdot 10^{-2}$ & $0.09 \pm 0.03$\\
    Pauli, lognormal & $(0.0 \pm 1.4) \cdot 10^{-2}$ & $0.30 \pm 0.03$ & $(-0.3 \pm 2.3) \cdot 10^{-2}$ & $0.10 \pm 0.03$& $(1.4 \pm 8.1) \cdot 10^{-2}$ & $0.04 \pm 0.02$\\  
    \hline
    \multicolumn{7}{|c|}{$n=12$} \\
    \hline
    Depolarizing, uniform & $(0.2 \pm 1.6) \cdot 10^{-2}$ & $0.57 \pm 0.03$ & $(0.2 \pm 1.7) \cdot 10^{-2}$ & $0.18 \pm 0.03$& $0.006 \pm 0.153$ & $0.08 \pm 0.01$\\
    Depolarizing, lognormal & $(0.2 \pm 1.4) \cdot 10^{-2}$ & $0.31 \pm 0.03$ & $(0.2 \pm 1.7) \cdot 10^{-2}$ & $0.09 \pm 0.02$& $-0.04 \pm 0.11$ & $0.06 \pm 0.01$\\
    Pauli, uniform & $(-0.4 \pm 2.0) \cdot 10^{-2}$ & $0.49 \pm 0.04$ & $(-0.8 \pm 3.5) \cdot 10^{-2}$ & $0.14 \pm 0.04$& $0.12 \pm 0.48$ & $0.09 \pm 0.03$\\
    Pauli, lognormal & $(0.1 \pm 2.7) \cdot 10^{-2}$ & $0.25 \pm 0.03$ & $(-0.2 \pm 4.2) \cdot 10^{-2}$ & $0.27 \pm 0.03$& $0.06 \pm 0.43$ & $0.04 \pm 0.01$\\  
    \hline
    \end{tabular}
\end{table*}}

\subsection{Comparison with existing ZNE techniques}

State of the art zero-noise extrapolation techniques rely on increasing the noise by adding extra gates or by implementing the existing ones in a different manner. Up to minor differences, we identified the following techniques in the literature, which can be directly compared to the proposed protocol:

\begin{enumerate}
    \item Unitary folding~\cite{giurgica-tiron_digital_2020}: parts of the circuit are repeated together with their inverses so that the length of the circuit is increased, while preserving the total unitary that would be implemented in the absence of noise. This can be done at the level of the entire circuit or individual gates and layers. In the latter case it is referred to as gate folding~\cite{he_resource_2020, dumitrescu_cloud_2018}.
    \item Pauli twirling and Pauli gate insertion~\cite{li2017_a}: with Pauli twirling, any local noise channel is transformed into a stochastic Pauli channel. Afterwars, stochastic Pauli noise is amplified by randomly adding Pauli gates.
\end{enumerate}

\begin{table*}[]
    \centering
        \caption{Extrapolation error for Hamiltonians of $n=6$ qubits with different ZNE techniques. The circuit depth was the same as in Fig. \ref{fig:epsilons_and_fits}. The folding extrapolations were obtained from linear extrapolation over 4 points.}
    \begin{tabular}{|c|c|c|c|c|}
    \hline
       Setup  & Pauli twirling & Unitary folding (gates) & Unitary folding (circuit) & This work \\
       \hline
       Ising A, depolarizing lognormal & $-0.027 \pm 0.030$ & $(4.0 \pm 4.4) \cdot 10^{-7}$ & $-0.036 \pm 0.078$ 
       & 
       $(-1.4 \pm 26.3) \cdot 10^{-4}$\\
       Ising B, depolarizing lognormal & $-0.16 \pm 0.19$ & $(5.0 \pm 4.6) \cdot 10^{-6}$ & $-0.12 \pm 0.03$ &
       $(-0.2 \pm 1.0) \cdot 10^{-2}$\\
       $\mathrm{H_2 O}$, depolarizing lognormal &
       $-0.028 \pm 0.055$
 & $(7.4 \pm 6.4) \cdot 10^{-8}$ & $-0.015 \pm 0.004$ & $(0.2 \pm 1.2) \cdot 10^{-3}$ \\
       \hline
    \end{tabular}
    \label{tab:sota}
\end{table*}

To compare the techniques, we took the depolarizing noise with lognormal error rate distribution and implemented the different ZNE techniques. As earlier, the ZNE error is averaged over 20 random realizations of noise. The results are shown in Table~\ref{tab:sota}. Our technique shows better results than circuit-level unitary folding and Pauli twirling, while looses to the gate-level unitary folding. 
We note here that unitary folding necessitates longer circuits, which naturally increase gate count and execution time, making it more vulnerable to cross-talk errors and limited coherence time. Therefore, we conclude that our protocol produces at worst comparable results to the state of the art techniques, without suffering from some of their drawbacks.

\section{Conclusions}
\label{sec:conclude}

Existing studies on the behavior of VQE-estimated energy with respect to noise primarily focus on homogeneous error models~\cite{fontana_evaluating_2021, dalton2022, kattemolle2022ability, rabinovich2024gate}. While a linear dependence between energy and the noise strength is well-established in this case, it does not automatically translate to the case of inhomogeneous noise. 
Motivated by the fact that gate errors are inhomogeneously distributed across pairs of qubits in a physical hardware, we propose a new method for zero-noise extrapolation.~In particular, we showed that by changing the abstract-to-physical qubit mapping, it is possible to vary the level of noise in a quantum circuit in a controllable way. This enables us to execute ZNE in an experimentally amenable manner.

While the proposed ZNE protocol can be applied to estimate noiseless expectation value of any observable with respect to any quantum circuit, for the purpose of demonstration we apply this method to VQE.
We found that the energy approximated by a noisy VQE circuit is approximately linear with respect to the circuit error sum. An analytic bound was derived to quantify the quality of the linear approximation. We found that this bound depends on the nature of the problem Hamiltonian and the error rates.  We numerically demonstrated that with the proposed ZNE protocol one can approximate the noiseless VQE energy with high accuracy, reducing the noise-induced error by approximately two orders of magnitude.
This was verified for a range of problem Hamiltonians, noise models and error distributions. 
Moreover, we proved that for a specific class of ansatze the proposed protocol recovers the exact noise free VQE energy. 

Next we examined how the performance of our proposed protocol gets impacted when considering permutation pools of limited sizes for the extrapolation; indeed considering all permutations maybe unfeasible especially for larger system sizes. We show that the ZNE extrapolation error is smaller than all energy scales of the problem, even when extrapolating over less than $10\%$ of all possible permutations. Furthermore, we demonstrated that while increasing the system size (while keeping the size of the permutation pool fixed) does induce some instability in the ZNE protocol, it can be removed with at most polynomially many permutations considered for extrapolation.

Finally, we compared the performance of the proposed protocol to the state of the art ZNE techniques. We showed that the performance of our hardware inspired ZNE, in the worst case, is comparable to the state of the art techniques, while not suffering from some of their drawbacks.

\acknowledgments

We thank Lianna Akopyan for stimulating discussions. 
O.L. and K.L. acknowledge the framework of the Roadmap for Quantum Computing (Contracts No.~868-1.3-15/15-2021 and No.~R2163). 
The code and the data produced in the work are available upon a reasonable request.



\appendix

\section{Relative deviation from linear trend in Eq.~\eqref{eq:e_vs_q_linear_avg}}
\label{appen_0}

We begin our analysis with a simplified form of \eqref{eq:e_vs_q_linear_avg} for the sake of brevity:

\begin{equation}
\label{Appen_eq:data_0}
    y = \Delta x + \delta y,
\end{equation}

where $y = (E-E_0)$, $\Delta =  (A - E_0)$, $x = \sum_{(j,k) \in T} d q_{jk}$ and $\delta y = \sum_{(j,k) \in T} (\sum_{l \in [1,d]} \epsilon^l_{jk}) q_{jk} = \sum_{(j,k) \in T} \Tilde{\epsilon}_{jk} q_{jk}$. 
Taking into account that both $x$ and $\delta y$ are error rate dependent terms, one can consequently conclude that $\delta y$ represents a deviation from the linear dependence $y = \Delta x$.  Nevertheless, one can upper bound the deviation $\delta y$ between the actual function $y$ as per \eqref{Appen_eq:data_0} and the linear dependence $y = \Delta x$. To accomplish this we consider the relative deviation:

\begin{equation}
    \label{appen_eq:rel_dev}
    \frac{\vert \delta y \vert}{\Delta x} = \frac{\vert \sum_{(j,k) \in T} \Tilde{\epsilon}_{jk} q_{jk} \vert}{d (A - E_0) \sum_{(j,k) \in T} q_{jk}}
\end{equation}

One can readily establish a bound on this quantity as:

\begin{equation}
    \label{appen_eq:bd1}
    \frac{\vert \delta y \vert}{\Delta x} \leq \frac{ \text{max}\{\vert\Tilde{\epsilon}_{jk}\vert\}_{(j,k) \in T}}{d (A - E_0)} = B_1
\end{equation}

This bound is not necessarily tight; however, one can improve it by taking into account the distribution of the error rates themselves:

\begin{equation}
    \label{appen_eq:bd2}
    \frac{\vert \delta y \vert}{\Delta x} \leq \frac{\vert T \vert \text{max}\{\vert\Tilde{\epsilon}_{jk}\vert\}_{(j,k) \in T}  \text{Diam} \{q_{jk}\}_{(j,k) \in T}}{d (A - E_0) \sum_{(j,k) \in T} q_{jk}} = B_2.
\end{equation}

Here $\text{Diam} \{A\}$ is the diameter of a set $A$. Combining \eqref{appen_eq:bd1} and \eqref{appen_eq:bd2} we arrive at the final result:

\begin{equation}
    \label{appen_eq:bd1}
    \frac{\vert \delta y \vert}{\Delta x} \leq \text{min} \{B_1, B_2\}
\end{equation}

\section{The zero-noise extrapolation over all permutations is exact}
\label{intercept_bnd}

Consider the function:

\begin{equation}
\label{Appen_eq:data}
    y = \Delta x + \delta y.
\end{equation}

We note that \eqref{Appen_eq:data} maps to \eqref{eq:e_permute}
when $y = (E-E_0)$, $\Delta = (A - E_0)$, $x = \sum_{(j,k) \in T} d \bar{q}_{\pi (jk)}$ and $\delta y = \sum_{(j,k) \in T} (\sum_{l \in [1,d]} \epsilon^l_{jk}) \bar{q}_{\pi (jk)} = \sum_{(j,k) \in T} \Tilde{\epsilon}_{jk} \bar{q}_{\pi (jk)}$. In our approach of zero-noise extrapolation we fit a linear approximation of the form $y = \alpha + \beta x$ to the function in \eqref{Appen_eq:data}. The constants $\alpha$ and $\beta$ are inferred by the least squares method where they have standard expressions in terms of sample statistics:
\begin{align}
    \alpha &= \langle y \rangle - \beta \langle x \rangle \label{appen_eq:alpha} \\
    \beta &= \frac{\operatorname{Cov}(x, y)}{\operatorname{Var} x} \label{appen_eq:beta}.
\end{align}
Here the angular brackets denote averaging over all samples (in our case, over all abstract-to-physical qubit mappings labelled by $\pi \in S_n$).
Substituting \eqref{Appen_eq:data} and \eqref{appen_eq:beta} in \eqref{appen_eq:alpha}, we obtain:
\begin{equation}
\label{appen_eq:alpha_2}
    \alpha = - \langle x \rangle \frac{\operatorname{Cov}(x, \delta y)}{\operatorname{Var} x}.
\end{equation}

Recall that for an exact zero-noise extrapolation we would get $\alpha = 0$. In deriving \eqref{appen_eq:alpha_2} we have used the result $\langle \delta y \rangle = 0$. This follows from two facts:
\begin{enumerate}
    \item $\langle \bar{q}_{\pi (jk)} \rangle = \frac{1}{n!} \sum_{\pi \in S_n} \bar{q}_{\pi (jk)} = \bar{q}_{a}$, where $\bar{q}_{a}$ is the average error;
    \item $\sum_{(j,k) \in T} \Tilde{\epsilon}_{jk} = 0$ by definition.
\end{enumerate}

Now we consider the covariance $\operatorname{Cov}(x, \delta y)$ in more detail. Owing to the fact that $\langle \delta y \rangle = 0$ we get

\begin{equation}
\label{Appen_eq:avg_cov}
    \operatorname{Cov}(x, \delta y) = \langle x \delta y \rangle = \frac{1}{n!} \sum_{\substack{(j_1, k_1) \in T \\ (j_2, k_2) \in T}} \Tilde{\epsilon}_{j_1 k_1} \langle \bar{q}_{\pi (j_1 k_1)} \bar{q}_{\pi (j_2 k_2)} \rangle
\end{equation}



We now group the summands in \eqref{Appen_eq:avg_cov} such that (a) $j_2 = j_1, k_2 = k_1$ (b) $j_2 = j_1, k_2 \neq k_1$ or $j_2 \neq j_1, k_2 = k_1$ and (c) $j_2 \neq j_1, k_2 \neq k_1$. Denoting the average $\langle \bar{q}_{\pi (j_1 k_1)} \bar{q}_{\pi (j_2 k_2)} \rangle$ for the aforementioned cases to be $\kappa_2, \kappa_1$ and $\kappa_0$ respectively, we arrive at the expression:

\begin{equation}
     \operatorname{Cov}(x, \delta y) = \sum_{(j k) \in T} \Tilde{\epsilon}_{j k} \left[n_2^{(j k)} \kappa_2 + n_1^{(j k)} \kappa_1 + n_0^{(j k)} \kappa_0 \right].
\end{equation}
Here $n_2^{(j k)}$ is the number of two-qubit gates acting on the qubit pair $(j, k)$, $n_1^{(j k)}$ is the number of two-qubit gates acting either on the qubit $j$ or $k$ and $n_0^{(j k)}$ is the number of gates that does not act either on $j$ or $k$. If these three values are independent on $(j, k)$, the covariance $\operatorname{Cov}(x, \delta y)$ vanishes owing to the fact that  $\sum_{(j,k) \in T} \Tilde{\epsilon}_{jk} = 0$ implying that $\alpha = 0$, which is what we expected for an exact zero-noise extrapolation.

We note here that $n_2^{(j k)}$, $n_1^{(j k)}$, $n_0^{(j k)}$ are independent of $(j,k)$ only for specific ansatz circuits. To characterise the structure  of such ansatze completely we depict the ansatz circuit as a multigraph with vertices corresponding to qubits and edges corresponding to two-qubit gates. Indeed $n_2^{(jk)}$ is the count of gates acting on $(j, k)$, $n_1^{(jk)} = \deg(j) + \deg(k) - n_2^{(jk)}$, and $n_0^{(jk)} = |T|d - n_1^{(jk)} - n_2^{(jk)}$. Thus, these values only depend on degrees of vertices and multiplicities of edges. This implies that the gate-independence condition can be satisfied by taking a regular graph (e.g.~a cycle) as the structure of the ansatz.

\bibliography{refs.bib}
\bibliographystyle{unsrt}

\end{document}